# Effect of long-range repulsive Coulomb interactions on packing structure of adhesive particles

Sheng Chen,[a] Shuiqing Li,[†a] Wenwei Liu,[a,b] Hernán A. Makse[b]

The packing of charged micron-sized particles was investigated using discrete element simulations based on adhesive contact dynamic model. The formation process and the final obtained structures of ballistic packings are studied to show the effect of interparticle Coulomb force. It was found that increasing the charge on particles causes a remarkable decrease of the packing volume fraction $\phi$ and the average coordination number $Z$, indicating a looser and chainlike structure. Force-scaling analysis shows that the long-range Coulomb interaction changes packing structures through its influence on particle inertia before they are bonded into the force networks. Once contact networks are formed, the expansion effect caused by repulsive Coulomb forces are dominated by short-range adhesion. Based on abundant results from simulations, a dimensionless adhesion parameter $Ad^*$, which combines the effects of the particle inertia, the short-range adhesion and the long-range Coulomb interaction, is proposed and successfully scales the packing results for micron-sized particles within the latest derived adhesive loose packing (ALP) regime. The structural properties of our packings follow well the recent theoretical prediction which is described by an ensemble approach based on a coarse-grained volume function, indicating some kind of universality in the low packing density regime of the phase diagram regardless of adhesion or particle charge. Based on the comprehensive consideration of the complicated inter-particle interactions, our findings provide insight into the roles of short-range adhesion and repulsive Coulomb force during packing formation and should be useful for further design of packings.

## 1 Introduction

Understanding the physics of spheres packings has both scientific and industrial importance since it has been linked to the microstructure and bulk properties of liquids, glasses, granular materials, as well as phase transition of colloidal systems.[1,2] Most previous studies have focused on two reproducible packing states for uniform spheres: random close packing (RCP) and random loose packing (RLP).[2-4] However, in real systems of nature or industry, complicated interactions among particles usually make packings deviate far from these two states.[5-10] For example, it has been demonstrated that adhesion can results in a decrease of overall packing fraction[8,11] whereas deformation of spheres under compression causes $\phi$ to increase up to 0.8.[10] To date, the relationship between the macroscopic packing structure and the microscopic interparticle forces is still open for research.

The primary concern of our work is the packing phenomena of micro-sized particles, which is ubiquitous in areas of material, astrophysics and environmental science.[7,12-14] For particles in the size range of 10 μm or smaller, the van der Waals (VDW) adhesion and electrostatic forces overcome the gravitational and frictional forces and become the dominant interactions that strongly affect packing structures.[7,15] The VDW adhesive force between micron-sized particles acts on length scales much smaller than the particle size, such that it is often regarded as a short-range interaction, which is quite different from the cases with nanoparticles or molecules regarding the VDW as a long-range force.[16] This strong short-range adhesion usually causes formation of particle agglomerates during packing process and hinders them from further compaction.[5] Previous studies have found that packing fraction of adhesive micro-particles varies in a range of $\phi = 0.165 \sim 0.622$ using a discrete element method (DEM).[9] Packings within $\phi = 0.20 \sim 0.55$ were also obtained for 4-5 μm particles both in experiments and simulations.[8] The latest work of Liu *et al.* combined the effects of particle inertia and interparticle adhesion, identifying a universal regime of adhesive loose packings (ALP) with packing fractions much smaller than RLP for particles across 1-100 μm.[11] Together with previous results from Ref.[17-21], a phase diagram in the $Z - \phi$ plane, derived in the spirit of Edwards' ensemble approach at the mean-field level, was presented for packings of frictionless, frictional, adhesive and adhesive-less spheres, as well as non-spherical particles. This phase diagram highlights that the universal packing regime resulted from adhesion can be described within a statistical mechanical framework. Based on these preliminary attempts, there remains a need for further investigation of the roles of long-range forces (e.g., electrostatic forces) on the packing state.

Compared with the van der Waals force, the electrostatic forces can exert their influence across a much longer distance. These long-range forces can cause profound changes in the structure of a granular flow, such as particle clustering, blockage as well as levitation, and offer the ability to manipulate particles at the microscales.[22-25] One of these electrokinetic phenomena related to packing of charged particles is electrophoretic deposition (EPD). It is derived from the transport of charged suspended particles under the influence of an external field and has been widely applied to the fabrication of wear resistant coatings as well as functional nanostructured films for electronic, biomedical and electrochemical applications.[14,26] Agreements have been reached that both the interparticle electrical interaction and the applied field significantly affect the quality of the deposits.[27,28] In colloidal suspensions, bulk phase transition and pattern formation were reported due to dipolar interactions under the action of a uniform ac field.[29] The suspension microstructure was found to be governed by the applied field strength.[30] In addition, particular attention was drawn to the problem of interstellar dust particles, ranging from the influence of charged particles on human exploration on the moon and Mars[12,31,32] to the coagulation among charged particles in protoplanetary disks.[33-35] Investigations of packing related to charged particle is of importance to develop deeper understanding of these processes. Nonetheless, few publications deal with that by now.

[a] *Key Laboratory for Thermal Science and Power Engineering of Ministry of Education, Department of Thermal Engineering, Tsinghua University, Beijing 100084, China. Email: lishuiqing@tsinghua.edu.cn.*
[b] *Levich Institute and Physics Department, City College of New York, New York 10031, USA. Email: hmakse@lev.ccny.cuny.edu*

To understand the complex, collective behaviour of particles during packing process and to further design, control or optimize the packing structure, people need to bridge the gap between the microscopic interparticle forces and the macroscopic packing structure. Numerical simulations by means of discrete element method (DEM) offer a helpful tool to understand packing of charged particles from a dynamic level where interparticle forces are explicitly considered. Focusing on properties of jammed configurations, traditional algorithms usually construct amorphous packings through either increasing the particle diameter at a given rate[36,37] or minimizing the energy repeatedly with increasing packing density[38]. Compared with these algorithms, DEM can readily incorporates much more complex interparticle forces and generate packings that are more comparable with practical physical systems. Generally, the forces in DEM includes short-range contact forces (e.g., elastic forces caused by particle deformation, sliding, twisting and rolling frictions caused by the relative motion of contact particles) and long-range forces (mainly electrostatic forces). This type of numerical simulations has been used both in packings of polydisperse particles[38] and packings of particles subjected to external electrostatic or gravitational fields.[6,39] However, to our knowledge, DEM investigation on packings of charged micron-sized particles is still limited by both proper consideration of highly-coupled contact forces and time-consuming pair-wise calculations of electrostatic forces.

Recently, a three-dimensional DEM for adhesive small particles based on the JKR (Johnson, Kendall and Roberts) model was developed by Li and Marshall,[40,41] and has been successfully applied to dynamic simulation of micro-particle deposition on both flat and cylindrical surfaces with a series of experimental validations.[13,15] Through the JKR model, the effect of VDW adhesion on the elastic deformation during contact of particles is properly described. Liu *et al*. incorporated a fast multipole method (FMM) into the DEM framework and achieved significant speedup for computation of the electric field induced by charged particles,[42,43] providing acceptable computational cost with a prescribe accuracy.

In this paper, this novel adhesive DEM is extended to ballistic packings of micron-sized charged particles, with the aim of elucidating the effect of interparticle Coulomb interaction on the packing structure. We also try to extract simple but effective rules that can predict packing properties, through extensive simulations and in-depth scaling analysis. In particular, the fluid effect is filtered out by assuming a vacuum condition to develop an "ideal" system. The structure of this paper is as follows: the computational set-up and a brief description of DEM framework are given in Sec. 2. Then we present the effect of Coulomb interaction on packing structures in terms of volume fraction $\phi$, coordination number $Z$ and radial distribution function $g(r)$. Deeper discussions are presented in Sec. 4, which include a force scaling analysis, a derivation of the scaling parameter and the packing state on a phase diagram. Finally, our conclusions are drawn in Sec. 5.

## 2 Models and Methods

### 2.1 Simulation conditions

We consider a random free falling of 2,000 spheres in.. direction with an initial injection velocity $U_0$ from a specific height $H$. The square plane for particle deposition in the bottom has a width of $L = 28 r_p$ ($r_p$ is the particle radius), which is set after a sensitivity analysis of $L$.[11] Periodic boundary conditions are applied along the horizontal, $y$ and $z$, directions to avoid lateral wall effect. The physical parameters of our simulations are listed in Table 1. Note that the surface energy $\gamma$ within the range 10-15 mJ/m$^2$, which is the typical range for $\gamma$ of silica microspheres, is used to reflect the effect of van der Waals adhesion.[7] The number of elementary charge $e_0$, which equals $1.6 \times 10^{-19}$ C, on a particle is determined according to the typical surface charge density due to diffusion and field charging for micro-particles,[7] and that of dust grains in astrophysical environments.[44] Due to the low conductivity among dielectric particles, the charge on a particle is assumed to be unchanged during the packing process. Higher-order multipoles, e.g., dipoles or quadrupoles, decay sufficiently fast with the distance and are ignored in this work. Such interactions, which may have effects on particles dynamics in the presence of electrodes or strong external fields,[22,30] will be left to future work.

**Table 1.** Parameters for simulation.

| Properties | Value | Unit |
|---|---|---|
| PARTICLE | | |
| Particle radius, $r_p$ | 1.0, 2.0, 5.0 | $\mu m$ |
| Density, $\rho_p$ | 2500 | $kg/m^3$ |
| Poison's ratio, $\sigma$ | 0.33 | - |
| Elastic modulus, $E$ | $2 \times 10^8$ | $Pa$ |
| Restitution coefficient, $e$ | 0.7 | - |
| Friction coefficient, $\mu$ | 0.3 | - |
| Surface energy, $\gamma$ | 10, 15 | $mJ/m^2$ |
| Charge on particles, $q$ | 0 ~ 500 | $e_0$ |
| TYPICAL PARAMETERS | | |
| Length, $L$ | 28 | $r_p$ |
| Hight, $H$ | 160 | $r_p$ |
| Initial velocity, $U_0$ | 0.5, 1.0 | $m/s$ |
| Particle number, $N_{tot}$ | 2000 | - |

## 2.2 Description of DEM approach

In order to elucidate the effect of particle-particle interactions on the packing structure, we simulate the packing process using DEM, simultaneously solving Newton's equations of translational and rotational motions for all particles. In our simulations, the long-range Coulomb force and short-range contact forces acting on each particle are taken into account as summarized in Fig. 1.

In the current work, the JKR model together with a dynamic damping model are employed to describe the contact force in the normal direction. The normal force acting on particle $i$ during a collision with particle $j$ can be expressed by

$$F_{ij}^n = F_{ij}^{ne} + F_{ij}^{nd} = -4F_C(\hat{a}_{ij}^3 - \hat{a}_{ij}^{3/2}) - \eta_N \mathbf{v}_{ij} \cdot \mathbf{n}. \quad (1)$$

The first term in Eq. 1 is derived from the JKR model which combines the effect of van der Waals adhesion and elastic deformation of particles. $F_C = 3\pi R_{ij}\gamma$ is the critical pull-off force and $\hat{a}_{ij}$ equals the contact radius $a_{ij}$ normalized by its value $a_0$ in zero-load equilibrium state, where the elastic repulsion is balanced by the adhesive attraction. Once given the deformation of particles $\delta_{ij} = r_i + r_j - |\mathbf{x}_i - \mathbf{x}_j|$, $\hat{a}_{ij}$ can be obtained by solving the equation

$$\delta_{ij}/\delta_C = 6^{1/3}\left(2\hat{a}_{ij}^3 - 4\hat{a}_{ij}^{1/2}/3\right) \quad (2)$$

In this equation, $\delta_C$ is the critical overlap at the pull-off point. For details see.[7] The second term of Eq. 1 represents the solid dissipation force which is proportional to the deformation rate $\mathbf{v}_{ij} \cdot \mathbf{n}$. The normal dissipation coefficient $\eta_N = \alpha\sqrt{m^*k}$ is described in literature[41] and the coefficient $\alpha$ is related to the restitution coefficient $e$ by a six order formula $\alpha = 1.2728 - 4.2783e + 11.087e^2 - 22.348e^3 + 27.467e^4 - 18.022e^5 + 4.8218e^6$. This is equivalent to the choice of a constant restitution coefficient $e$ in inertia-dominant collisions. In the presence of adhesion the apparent restitution coefficient in a collision ranges from 0 for low-velocity sticking cases to the constant value $e$ for inertia-dominant collisions.[45,46]

Besides the deformation in the normal direction, we use spring-dashpot-slider models to calculate interparticle sliding,

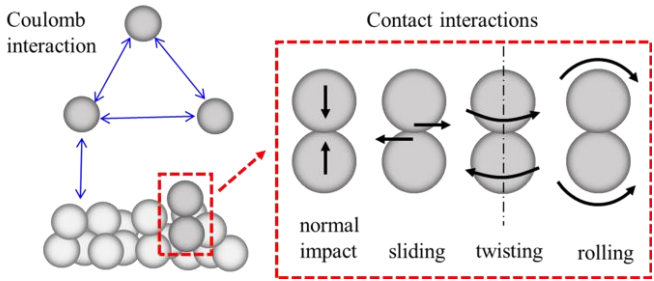

Fig 1. Schematic representation of the long-range Coulomb interaction and short-range contact interactions.

twisting and rolling frictions.[7,41] The force and torques are written as

$$F_{ij}^s = -\min\left(k_T(\int_{t_0}^t \mathbf{v}_{ij}(\tau)\cdot\mathbf{t}_S d\tau) + \eta_T \mathbf{v}_{ij}\cdot\mathbf{t}_S, F_{ij,crit}^s\right),$$
$$M_{ij}^t = -\min\left(\frac{k_T a^2}{2}\int_{t_0}^t \Omega_T(\tau)d\tau + \frac{\eta_T a^2}{2}\Omega_T, M_{ij,crit}^t\right), \quad (3)$$
$$M_{ij}^r = -\min\left(4F_C(a/a_0)^{3/2}\cdot(\int_{t_0}^t \mathbf{v}_L(\tau)d\tau)\cdot\mathbf{t}_R, M_{ij,crit}^r\right),$$

where $\mathbf{v}_{ij}(\tau)\cdot\mathbf{t}_S$, $\Omega_T$ and $\mathbf{v}_L$ stand for the relative sliding velocity, twisting velocity and rolling velocity between two contact particle. When these resistances reach certain critical limits, termed as $F_{ij,crit}^s$, $M_{ij,crit}^t$ and $M_{ij,crit}^r$, they stay constant and the particles start to slide, spin or roll irreversibly relative to each other. These critical limits are all related to the effect of van der Waals adhesion and can be expressed as

$$F_{ij,crit}^s = \mu_f \left|F_{ij}^{ne} + 2F_C\right|,$$
$$M_{ij,crit}^t = 3\pi a_{ij} F_{ij,crit}^s / 16, \quad (4)$$
$$M_{ij,crit}^r = -4F_C \hat{a}_{ij}^{3/2}\theta_{crit} R_{ij}.$$

The friction coefficient $\mu_f$ is set to be 0.3, and for the critical rolling angle we use $\theta_{crit} = 0.01$, which are set based on experimental data.[13,15,47] Our adhesive 3D DEM has been successfully applied to simulations of various adhesive particle behaviors, including particle-wall collisions[15] and deposition of particles on a fiber.[13]

Besides the aforementioned short-range contact forces, the presence of charged particles induces an electric field, which decays slowly with distance away from each particles. And the charged particles in turn bare the force exerted by the induced field in the form of

$$\mathbf{F}_{E,i} = q_i \mathbf{E} = q_i \sum_{j\neq i}\frac{q_j \mathbf{r}_{ij}}{4\pi\varepsilon_0 r_{ij}^3}, \quad (5)$$

where $q_i$ is the charge on particle $i$, $\mathbf{r}_{ij} = \mathbf{x}_i - \mathbf{x}_j$ is the vector from the centroid of source particle $j$ to the target one and $\varepsilon_0$ is the permittivity of the vacuum. The long-range feature of the electrostatic forces poses challenges in calculating the pair interactions among thousands of particles. For a system with $N$ particles, the cost of direct calculation of the pair-wise Coulomb interactions scales as $O(N^2)$ leading to an unacceptably low simulation efficiency for systems with large $N$. This difficulty can be overcome by employing a fast multipole method which obtains an approximation for electrostatic forces on a target particle exerted by a group of particles located sufficiently far away. Particles are separated into boxes and the electric field generated by box $l$ can be expressed in terms of the multipole expansion as

$$\mathbf{E}(\mathbf{r}) = \sum_{m=0}^{+\infty}\sum_{n=0}^{+\infty}\sum_{k=0}^{+\infty}\frac{(-1)^{m+n+k}}{m!n!k!}I_{l,mnk}\frac{\partial^{m+n+k}}{\partial x^m \partial y^n \partial z^k}\mathbf{K}(\mathbf{r}-\mathbf{r}_l), \quad (6)$$

where $I_{l,mnk}$ is the box moment and the interaction kernel $\mathbf{K}(\Delta \mathbf{r}) = \Delta \mathbf{r} / 4\pi\varepsilon(\Delta r)^3$ depends only on the location of box centroid $\mathbf{r}_l$ and the target point $\mathbf{r}$. The computing cost thus is reduced to $O(N \log N)$, with the precision controlled by an analytic error bound.[48] For details see.[7,43] The periodic images in the virtual domains which are far away from the physical domain are approximated by uniformly distributed charges with enough high precision, as given in the Appendix. Compared with direct calculation of each periodic images, this average-field method is less time-consuming and has a good applicability for simulation of packing systems.

## 3 Effects of Coulomb interaction on packing structure

This section presents the results of our DEM simulations. We analyze structure of packings obtained with different charge on particles in terms of the most commonly used concepts, such as volume fraction $\phi$, coordination number $Z$ and radial distribution function $g(r)$.

As shown in Fig. 2, from the macro perspective, the charge on particles significantly affects the final packing structure. With other parameters fixed, a higher charge of particles will lead to a looser packing structure. The expansion effect is further quantified in Fig. 3, where the variation of volume fraction $\phi$ as a function of particle charge is plotted for three typical series of packings. The volume fraction for packings of neutral particles $\phi_0$ ranges from 0.270 to 0.363 here, which still lies in the range of adhesive loose packings of non-charged particles reported in the literatures.[9,11] As particle's charge $q$ increases, $\phi$ starts to decrease lightly and then rapidly drops as $q$ further increases. We redraw the data in the form of $\phi_0 / \phi - 1$, which is regarded as the relative expansion of packed beds, in double logarithmic coordinates. It can be found that, a line with a slope of two nicely describe the variation tendency of $\phi_0 / \phi - 1$, implying that the expansion effect has a relation to the inverse square law of Coulomb interaction. In these cases, the relative decrease of $\phi$ can reach a maximum of 40%. Regarding the charge on particles as a controllable parameter, this expansion effect suggests the ability of tuning the interparticle interactions and manipulating the packing structures. It should, however, be noted that the maximum value of $q$ is limited by both the charging mechanisms[7,44] and the onset of particle levitations where incident particles are repelled away from deposited particles before reaching the packed bed.[23] Therefore, under the condition of a sufficiently large $q$, the square law will breakdown. Packing of particles with charges beyond this limit is not within the scope of this study.

In addition to $\phi$, we consider the radial distribution functions (RDF), $g(r)$, to gain some more insight into the microstructure of packings. The RDF is defined as the probability of finding a particle at a given distance $r$ from a reference one and indicates structural changes when shoulders or peaks appear. A plot of the RDF for packings of both charged and uncharged particles is shown in Fig. 4. The sharp peak observed at $r = 2r_p$ corresponds to the first contact shell of particles that are in touch with the reference particle. The general trends of these two RDFs, including the absence of any peaks after $r/2r_p = 2.0$ and a narrow first peak, are parallel to the existing results for micron-sized neutral particles.[9] We further indicate that Coulomb interaction does not bring detectable crystallization based on the observation of the absence of a peak at $r/2r_p = \sqrt{2}$ or $\sqrt{5}$, which are typical of crystal packings.[49] The most remarkable difference between these two RDFs is that the peak at $r/2r_p = \sqrt{3}$ which relates to the configuration of edge-sharing equilateral triangles vanishes for charged packing. This change suggests that the interparticle Coulomb interaction makes the particles tend to form straight chains rather than compact cells resulting in a looser structure, which supports the decrease of $\phi$ discussed above.

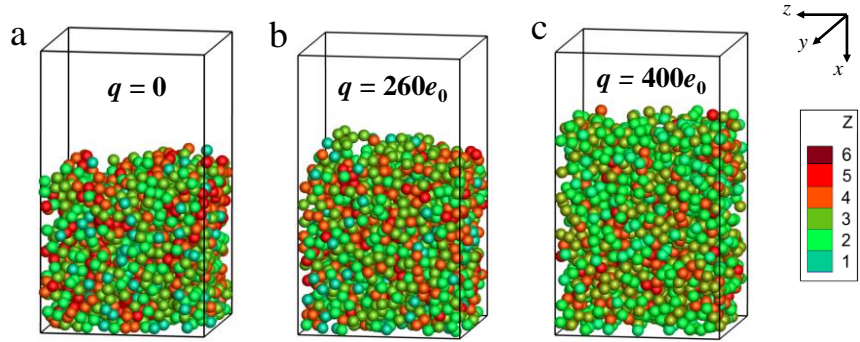

Fig. 2. Typical packing structures for $r_P$ = 2.0 μm $U_0$ = 1.0 m/s with (a) $q$ = 0, (b) $q$ = 260$e_0$ and (c) $q$ = 400$e_0$. Different colours stand for different coordination number $Z$

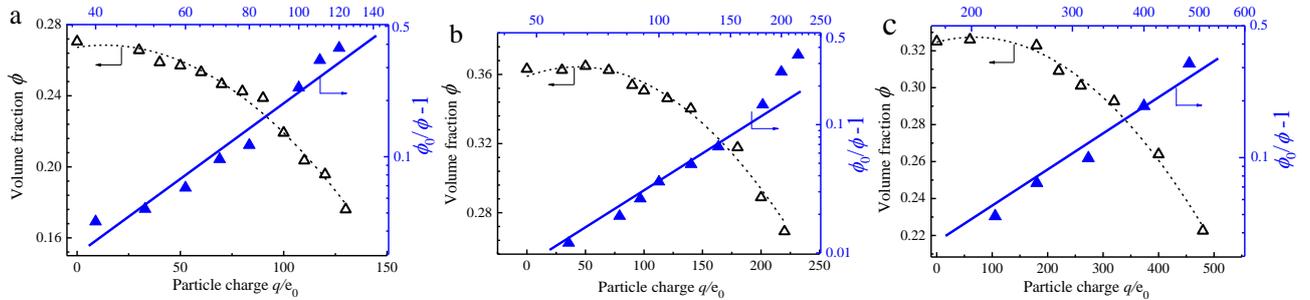

Fig. 3. Volume fraction as a function of particle charge for (a) $r_p$ = 1.0 μm, $U_0$ = 1.0 m/s, (b) $r_p$ = 1.0 μm, $U_0$ = 2.0 m/s and (c) $r_p$ = 2.0 μm, $U_0$ = 1.0 m/s.

We also observe the contact condition of particles inside the packings. In the present study, the concept of contact refers to geometrical contact which may include 'trivial' contacts with zero force.[19] Particles inside the packing are in contact with each other to maintain the stability of the structure. And the coordination number $Z$, defined as the number of contact of a particle in the packing, is another observable measurement of packing structure. A quantitative description of the contact condition is given in the form of coordination number distribution $f(Z)$ as shown in Fig. 5. It can be seen that, with other parameters fixed, the mean value of the distributions moves to the left with the increasing of particle charge. The populations of $Z = 4 \sim 6$ on the right of the peaks shrink with higher $q$ while the $Z = 1 \sim 2$ populations grow, indicating that chain-like agglomerates are more likely to be formed inside. This chain-like network acts as a skeleton to support the highly porous structure. Interestingly, decreasing the particle radius (compare Fig. 5a and Fig. 5c) or decreasing the injection velocity (compare Fig. 5a and Fig. 5b) has a similar effect as the interparticle Coulomb interaction, i.e., makes the distribution move toward left. This kind of similarity leads us to find a general principle that can bring together the role of interparticle interactions at different ranges (e.g., the short-range contact interactions and the long-range Coulomb interaction) and quantify the combined effect of particle size, velocity, surface energy and charge. This principle is discussed next.

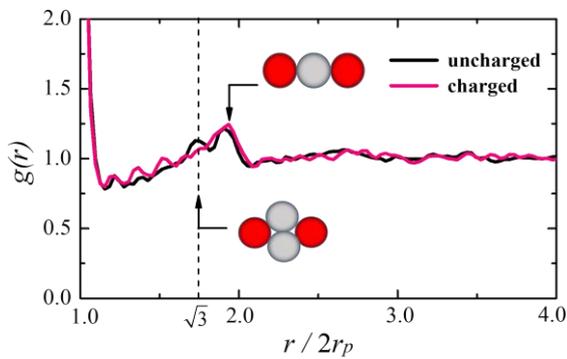

Fig. 4. Radial distribution function for the packings of neutral and charged particles, corresponding to the packing conditions of (a) and (c) in Fig. 2.

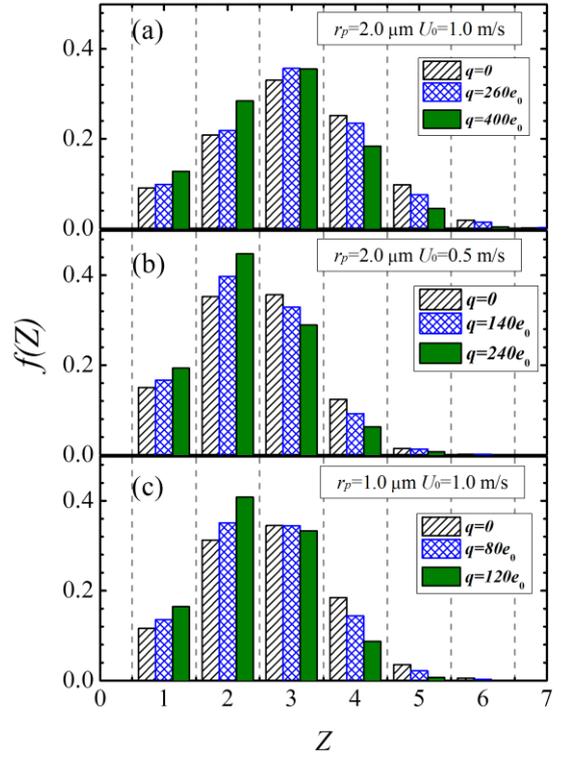

Fig. 5. Effect of particle charge on coordination number distributions. (a) stands for packing with $r_p$ = 2.0 μm $U_0$ = 1.0 m/s, (b) stands for $r_p$ = 2.0 μm $U_0$ = 0.5 m/s and (c) stands for $r_p$ = 1.0 μm $U_0$ = 1.0 m/s.

## 4 Discussion

Despite a clear description of the packing structures obtained under different conditions, we still need a general criterion to measure the impacts of different factors on the structure. This criterion, if available, should be related to the forces governing particle motions. To make progress with this aspect, we shift our focus to the forces that control the packing structural evolution.

### 4.1. Force scaling analysis and force distribution

For the system considered here, the forces exerted on particles can be classified into two categories according to their effective ranges: the short-range contact forces and the long-range interaction forces. The former includes normal adhesive force $F_{ne}$, damping force $F_{nd}$ and resistances in tangential direction. It should be noted in particular that the interparticle van der Waals force decays quickly with the distance and has been integrated into the JKR model together with the normal elastic force. Thus, it is classified into the category of short-range contact forces. The second category only includes the Coulomb interaction in the present study.

Intuitively, the repulsive Coulomb force between deposited particles may bring in an "expansion" effect to the packed bed. However, no detectable change of the packing structure was found when we changed the charge on the particle from 0 to $400e_0$ (corresponding to the charging states in Fig. 2) after the packing has been formed (see Table 2). The reason is that once the packing has been formed, all the particles inside the packing are bonded into the contact network and the strength of the repulsive Coulomb force is much weaker than the dominant van der Waals adhesion. This can also be inferred from the scaling analysis which is characterized by the ratio of van der Waals adhesion force to Coulomb electrostatic force. Here the typical van der Waals force can be represented by the pull-off force $F_C = 3\pi\gamma R$ while the Coulomb force is $F_E = q^2 / 4\pi\varepsilon_0 r_p^2$. With parameters listed in Table 2, we evaluate the maximum Coulomb force and have $F_C / F_E \approx 10^4 >> 1$ which confirms of the statement above.

Besides the macroscopic structure parameters presented in Table 2, it is instructive to present the forces carried by the contact networks with a visualization of repulsive and attractive normal forces, as shown in Fig. 6. For clarity, the samples were just slices with a width of $3r_p$ taken from the packings. Such force patterns are typical for packing of particles in the presences of adhesion.[50] Introducing the repulsive Coulomb interaction into the equilibrated samples, only a small fraction of normal forces changes from repulsive ones to attractive ones and the main part of the force network remains essentially unchanged. To characterize such force patterns in a more quantitative way, the force distributions of these two samples are plotted in Fig. 7. The forces are normalized by the typical pull-off force $F_C$, which is the criterion for contact break-up. The distributions are narrow (within the range of $|F^n / F_C| < 0.1$) and almost symmetric, which can be directly inferred from the force balance between the repulsive elastic force and the attractive adhesion force. After introducing the Coulomb force, a shift of the distribution from repulsive (positive) to attractive (negative) to adaptively balance the applied repulsive Coulomb force. This change of force distribution further indicates that, the repulsive Coulomb interaction is too weak to cause any break-up of contact pairs or rearrangement of packing structures after they are formed, provided $\Delta F^n / F_C << 1$.

**Table 2.** Packing structures when charged after formation.

| Particles charge ($e_0$) | Volume fraction | Coordination number |
|---|---|---|
| 0 | 0.3249 | 3.2056 |
| 260 | 0.3248 | 3.2056 |
| 400 | 0.3247 | 3.2056 |

However, due to the long-range characteristic of the Coulomb interaction, it still plays an important role before particles are bonded into the force network. Fig. 8 shows the evolution of the kinetic energy $E_k$ of an incident particle as a function of the distance between its centroid and the top of the packed bed. The kinetic energy and the distance have been normalized by the initial kinetic energy $E_{k0}$ and the particle radius $r_p$ respectively. The Coulomb force exerted by the deposited particles decelerates the incident one, leading to the energy conversion from kinetic form to Coulomb potential. Higher charge results in a lower impact velocity at the moment of contact as shown in Fig. 8. The relationship between particle inertia and the packing structure has been discussed for neutral particles with the conclusion that lower particle inertia results in a looser packing structure.[11] These results provide compelling evidence that, the long-range Coulomb interaction influences the packing structure indirectly through its influence on the inertia of particles *before* they are bonded into contact networks. These findings extend those of Liu *et al.*,[11] clarifying the acting range of Coulomb force during the packing of adhesive micron-sized particles, which highlights the original features brought out by our simulations.

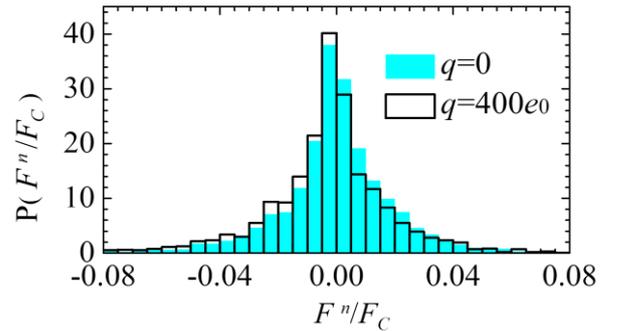

Fig. 7. Probability distribution functions of dimensionless normal force values for packings of neutral and charged particles.

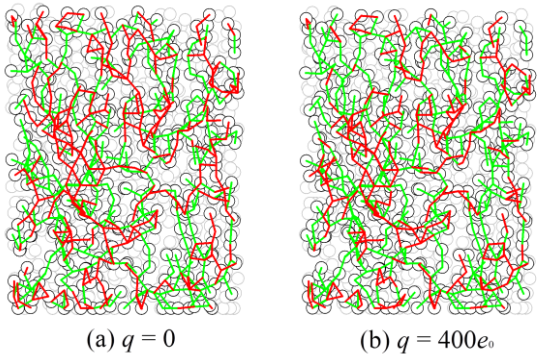

Fig. 6. Force-carrying structures in (a) neutral and (b) charged packing samples. For clarity, the samples were just slices with a width of $3r_p$ taken from the packings. Red and blue lines stand for compressive and tensile interactions respectively and the width of the lines is related to the magnitude of these forces.

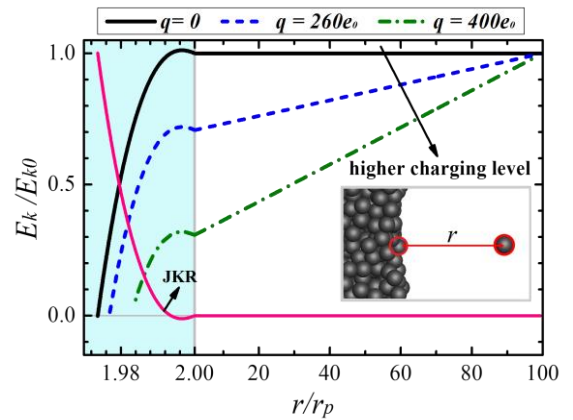

Fig. 8. The variation of the kinetic energy $E_k$ of an incident particle as a function of the distance between the particle and a deposited one. The vertical dash line divides the contact state and the non-contact state. The JKR line stands for the energy transferred from kinetic form to the sum of released surface energy and stored elastic energy.

### 4.2. A scaling parameter for ALP with Coulomb interaction

It has been found that, for packings of adhesive micron-sized particles, the volume fraction and the corresponding coordination number can be related to the ratio of interparticle adhesive force and the particle inertia.[11] The interparticle adhesion is characterized by the surface energy $\gamma$ through the JKR model and the particle inertia is a function of its size and velocity. A dimensionless adhesion parameter $Ad = \gamma/(\rho_P U_0^2 r_P)$, which combines all these parameters, has been successfully used to quantify the relative importance of the particle adhesion and inertia.[11,13,40] This parameter has been used in previous study of packing for neutral micro-particles, where a universal regime of adhesive loose packing for $Ad > 1$ is identified.[11] Taking the long-range Coulomb interaction into account, we modify the parameter $Ad$ by introducing a concept of effective impact velocity, which is written as

$$U_{eff}^2 = U_0^2 (1 - E^*). \quad (7)$$

Here, $U_0$ stands for the incident velocity of particles and the dimensionless charge parameter $E^* = K_{eff} q^2/(\varepsilon_0 U_0^2 r_P^4 \rho_P)$ is a measurement of the relative importance of the Coulomb interaction compared with the particle inertia. The coefficient $K_{eff}$ is related to the total particle number and the domain dimension, and is fixed at $K_{eff} = 20 N_{tot}(L/r_P)^{-2} = 51$ for all the cases in the present study. Then a modified adhesion parameter $Ad^*$ can be written as

$$Ad^* = \frac{\gamma}{\rho_P U_{eff}^2 r_P} = \frac{\gamma}{\rho_P U_0^2 (1-E^*) r_P}. \quad (8)$$

This parameter thus combines the effects of particle velocity $U_0$, size $r_p$, surface energy $\gamma$ and charge $q$.

We have simulated a series of packings with different $r_p$, $U_0$, $\gamma$ and $q$. The variation of the volume fraction $\phi$ and the average coordination number $<Z>$ are plotted as a function of $Ad^*$ respectively in Fig. 9. It can be found that both $\phi$ and $<Z>$ decrease monotonically as $Ad^*$ increases. At first, the data points are almost located around a straight line confirming the adhesion-controlled regime recently obtained in.[11] The decrease of $\phi$ and $<Z>$ can be attributed to the competition between particle's adhesion and inertia.

When particles are being packed, the VDW adhesion force tends to attract particles and make them stick together while the particle inertia will urge them to move and impact with other particles. If adhesion is stronger than particle inertia (corresponding to a large $Ad^*$), particles will be caught at the first moment of impact and hardly move or roll (termed as hit-and-stick phenomenon) so that a loose packing structure is easier to form.[51] With the increase of particle size or velocity, its inertia becomes stronger, leading to violent collisions and the particle bed will rearrange upon collisions to form a denser packing.

The long-range Coulomb interaction exerts its influence on a particle as soon as it enters the computational domain and continuously decelerates the particle. This effect causes an increase of $Ad^*$ and results in relatively loose packings. In these cases, the adhesion parameter $Ad^*$ successfully combines the effects of particle's properties and bridges the gap between the macroscopic packing structure and the microscopic interparticle forces. However, it should be noted that all the cases in the present study are located in the regime of $Ad^* > 1$, where the adhesion dominates the packing. When $Ad^* < 1$ the packing structure will be related to the interparticle friction,[52,53] and our scaling may break down since $Ad^*$ does not include the effect of friction.

Interestingly, as $Ad^*$ further increases, both $\phi$ and $<Z>$ deviate from the straight line and start to enter a prolonged plateau. This deviation is mainly due to the existence of an asymptotic adhesive loose packing (ALP) limit at $<Z> = 2$ and $\phi = 1/2^3$, which is an important conjecture in the latest packing studies.[11] This point is naturally related to the observation that $\phi = 1/2^d$ is the lower bound of saturated sphere packings in $d$ dimensions.[54] Moreover, $Z = 2$ is the minimal possible value of the coordination number for particles bonded into a network spanning the packing.

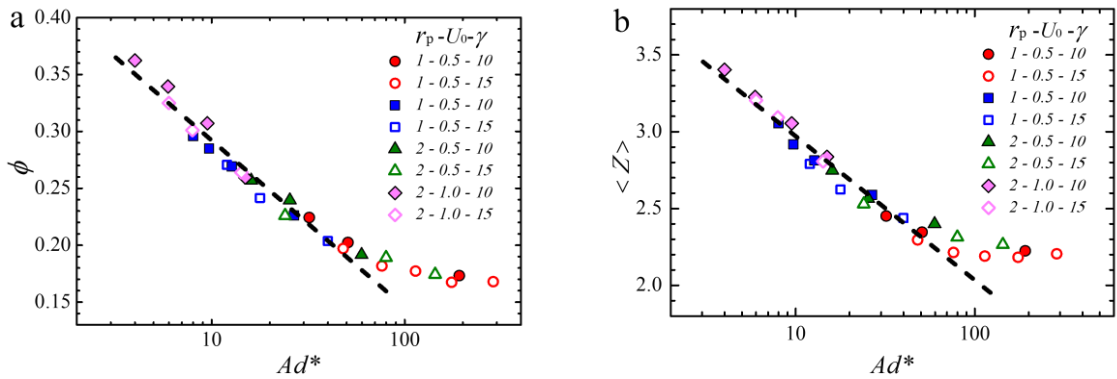

Fig. 9. Semi-log plot of (a) the packing volume fraction $\phi$ and (b) the average coordination number $<Z>$ as a function of the modified adhesion parameter $Ad^*$.

### 4.3. Phase diagram

Apart from the dynamic analysis above, we have also compared our simulation results with an analytical representation of the equation of state $\phi(Z)$ for adhesive packing which is developed based on the Edwards' ensemble approach.[11,17,18,20,21] The approach starts with the tessellation of the total volume of the packing: $V = \sum_{i=1}^{N} W_i$, where $W_i$ is the Voronoi volume of a reference particle $i$. The key step is to use a statistical mechanical description of the average Voronoi volume $\overline{W} = \langle W_i \rangle$, which implies that $V = N\overline{W}$ and the packing fraction becomes $\phi = V_0/\overline{W}$. Here $V_0$ is volume of a sphere with radius $r_p$ in the packing. In turn, $\overline{W}$ can be expressed in terms of the cumulative distribution function $P(c,Z)$, which is the integration of the pair distribution function $p(c,Z)$ for finding the boundary of the Voronoi volume at a distance c from the sphere centre. Then the average Voronoi volume $\overline{W}$ is

$$\overline{W}(Z) = V_0 + 4\pi \int_{r_p}^{\infty} c^2 P(c,Z)\, dc \;, \qquad (9)$$

where $p(c,Z) = -dP(c,Z)/dc$. For $P(c,Z)$ one can derive a Boltzmann-like form using a factorization assumption of the multi-particle correlation function into pair correlations to find

$$P(c,Z) = \exp\left\{-\rho \int_{\Omega(c)} d\mathbf{r}\, g_2(\mathbf{r},Z)\right\}. \qquad (10)$$

Here $\rho = N/V = 1/\overline{W}$ is the number density and $g_2(\mathbf{r},Z)$ is the pair correlation function of two spheres separated by $\mathbf{r}$. The volume $\Omega(c)$ is an excluded volume for the $N$-1 spheres outside of the reference sphere, since otherwise they would contribute a Voronoi boundary smaller than $c$. The exponential form of Eq. 10 is the key assumption of the mean-field approach. To capture the substantial correlation between each particle and its neighbours in the packing, the pdf $g_2(\mathbf{r},Z)$ needs to be well modelled. The function of has been discussed in detail in[11] with four distinct contributions following the results of simulations and mean-field models of metastable glasses, so that we are not going to address it again here. Solving Eq. 9, 10 and $g_2(\mathbf{r},Z)$ defined in[11] numerically for $\overline{W}$ we obtain the unique equation of state $\phi(Z)$ as shown in Fig. 10. In this adhesive loose packing regime, our simulation results are in substantially good agreement with the theory. This findings extend those of Liu *et al.*,[11] confirming that the addition of the interparticle Coulomb interaction does not break the microstructural feature of adhesive packings. Based on the discussion in Sec. 4.2, these interesting results are mainly caused by two reasons. One is that for packings in the regime of $Ad^*>1$, where adhesion still dominates over other interactions even if the Coulomb force exists, particles are stabilized by attractive adhesive forces as they are in.[11] The introduced Coulomb interaction makes few contributions to the equilibrium of the packings after particles get in contact with each other. On the other hand, during the formation of the packing, namely the falling process, the repulsive Coulomb force will decelerate the particles, resulting in a lower velocity thus a larger $Ad$. To correctly characterize the deceleration effect caused by Coulomb force which is not considered in the previously defined $Ad$, the modified $Ad^*$ is then put forward. As a consequence, the phase diagram of $<Z>$ versus $\phi$ will not be influenced by the introducing of Coulomb force, implying the universality in the low packing density regime regardless of adhesion or particle charge. As for the $\phi \sim Ad^*$ and $<Z> \sim Ad^*$, the increase of particle charge just leads to a shift to larger $Ad^*$ (low packing fraction) regime. Therefore the contact and bulk properties can be treated substantially the same as those in.[11] However, on the other hand, the tendency of moving toward the ALP point with increasing of particle charge in the phase-diagram, suggests that charging the particle is a natural way to increase the packing porosity in disordered arrangements.

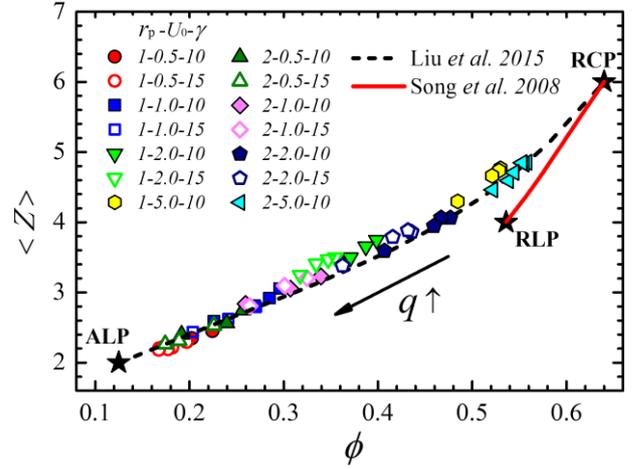

Fig. 10. Packing states on the phase diagram for charged mirco-sized particles: simulation data and theoretical prediction.

## 5 Conclusions

In this paper, we have presented a computational study of the packing of charged micron-sized particles. By using JKR-based adhesive contact models and a fast multipole method, we are able to make progress to bridge the gap between the macroscopic packing structure and the microscopic interparticle forces. The effect of long-range Coulomb force on packing structure is quantified in terms of volume fraction, coordination number and radial distribution function. And further analysis from both dynamic and statistical mechanical levels is conducted to clarify the connections and the differences between the effects of the short-range van der Waals force and the long-range Coulomb force. We found in our simulations that the presence of long-range Coulomb interaction results in a looser packing structure through its influence on particle inertia. The relative decrease of the volume fraction, recorded as $\phi_0/\phi - 1$, approximately follows a square law with the increase of particles charge up to a maximum of 40%. However, this effect is suppressed by short-range adhesion once particles are bonded into the contact network. Furthermore, a modified $Ad^*$ was derived to clarify the combined effects of particle inertia, adhesion and Coulomb interaction. With $Ad^*$ increasing, both the volume fraction $\phi$ and the average coordination number $<Z>$ decrease monotonously.

This suggests that $Ad^*$ can be used to predict and further design the macroscopic structure of packing of charged adhesive particles. We have also shown that the packing state of charged micron-sized particles can be well described by the latest derived adhesive loose packing (ALP) regime in the phase diagram[11] and increasing the charge of particle makes packing states move toward the ALP point. This indicates the universality of the analytical presentation for packings of adhesive or charged particles based on the Edwards' ensemble approach.

Several avenues for future investigation have also been indicated based on our results. First, our investigation is restricted to the specific packing problem of charged particles in the absence of external fields. Since there also exists ubiquitous phenomena of particle chaining and ordering in the presence of external field.[22,30,55] Expanding our model to include the effects of higher-order multipoles and external field, which may cause significant change of packing properties, seems worth pursuing. Moreover, further investigation is also needed to understand the nature of packings in the vicinity of ALP. It would be highly interesting to find out whether an ALP is indeed observed in a real physical system.

## Appendix: description of the average-field method

Consider our packing system of $N$ charged particles with a periodic boundary conditions mentioned previously. A periodic array of replicated systems is created as shown in Fig. 11. Because of the long-range nature of the Coulomb interaction, the force exerted on a particle includes contributions from the other particles inside the physical domain and all their replica images over the periodic array. The total field at location of the particle $i$ is

$$\mathbf{E}_i = \frac{1}{4\pi\varepsilon_0} \sum_{\mathbf{n}} \sum_{j=1}^{N} \frac{\mathbf{r}_{ij} + L\mathbf{n}}{\left|\mathbf{r}_{ij} + L\mathbf{n}\right|^3} \cdot q_j, \quad (11)$$

where $q_j$ is the charge on particle $j$ and $L$ is the cell dimension. The sum is over all integer vectors $\mathbf{n} = (n_y, n_z)$ with the term $i = j$ omitted when $|\mathbf{n}| = 0$. The whole periodic system is divided into two regions according to the distance from the central computational domain. Then we rewrite Eq. 11 as the summation of contributions from these two regions

$$\mathbf{E}_i = \mathbf{E}_{i,in} + \mathbf{E}_{i,out}$$
$$= \frac{1}{4\pi\varepsilon_0} \sum_{|n_y|,|n_z|\leq 1} \sum_{j=1}^{N} \frac{\mathbf{r}_{ij} + L\mathbf{n}}{\left|\mathbf{r}_{ij} + L\mathbf{n}\right|^3} \cdot q_j + \frac{1}{4\pi\varepsilon_0} \sum_{|n_y|or|n_z|>1} \sum_{j=1}^{N} \frac{\mathbf{r}_{ij} + L\mathbf{n}}{\left|\mathbf{r}_{ij} + L\mathbf{n}\right|^3} \cdot q_j \quad (12)$$

The first term on the right of Eq. 12 accounts for the contribution of particles inside the central computational domain and its neighboring virtual domains enclosed by the red square as shown in Fig. 11. As stated in previous, we use a multiple expansion method to calculate the electric field induced by this part of particles. On the other hand, the second term of Eq. 12 is not truncated and therefore extend over the entire system. For simplicity, these periodic images in the virtual domains outside the red square are approximated by uniformly distributed charges and the summation is thus turned into an integral, yields

$$\mathbf{E}_{i,out} = \frac{1}{4\pi\varepsilon_0} \sum_{|n_y|or|n_z|>1} \sum_{j=1}^{N} \frac{\mathbf{r}_{ij} + L\mathbf{n}}{\left|\mathbf{r}_{ij} + L\mathbf{n}\right|^3} \cdot q_j \approx \frac{1}{4\pi\varepsilon_0} \oint_{out} \frac{\mathbf{r}}{\left|\mathbf{r}\right|^3} \sigma dv. \quad (13)$$

The integral is taken over the region occupied by periodic images of packed particles outside the red square and the average charge density is written as $\sigma = \sum_{j=1}^{N} q_j / V_{bed}$, where $\sum_{j=1}^{N} q_j$ and $V_{bed} = L^2 \cdot H_p$ are the sum of charges and the volume of packed bed inside each domain respectively. This average-field method accelerates the computation of the field induced by charged particles over periodic arrays since the only quantity needed to be updated during computation process is the sum of particle charges and the integral can be determined without knowing the exact location of particles. The physical basis for this method is that when charged particles (the source) are sufficiently far from the target point, the charges on them can be redistributed over nearby regions without causing obvious changes in the field at the target point. For packing systems here, particles are uniformly distributed inside the domains and the charges on them thus can be well approximated by a homogeneous charge density $\sigma$. Despite the fact that further investigation on analytical error bound is still needed to evaluate this method, it achieves sufficient accuracy throughout our simulations for packing of charged particles.

## Acknowledgements

This work has been funded by the National Natural Science Funds of China (No. 51390491) and by the National Key Basic Research and Development Program (No. 2013CB228506). S. Q.

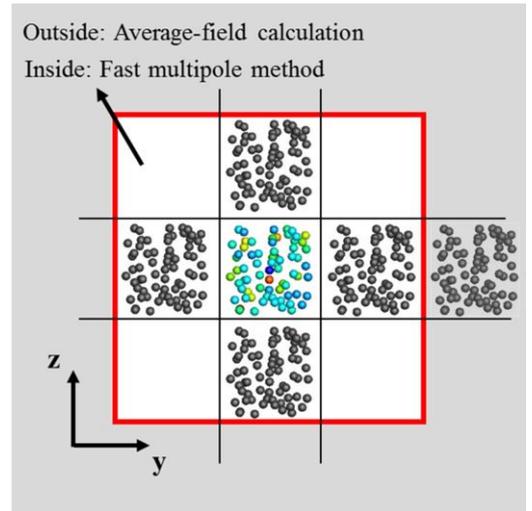

Fig. 11. Illustration of the periodic boundary condition. The electric field induced by particles inside the physical domain and its neighboring virtual domains (the region inside the red square) is calculated by a fast multipole expansion method. The contribution of particles outside the red square is approximated by an average field.

Li is grateful to Prof. Jeff Marshall at Vermont and Dr. G. Q. Liu at Tsinghua for helpful discussions. H. Makse acknowledges funding from DOE and NSF.